# Backward air lasing actions induced by femtosecond laser filamentation: influence of population inversion lifetime

**Hongqiang Xie[1, 2], Guihua Li[1], Wei chu[1, †], Bin Zeng[1], Jinping Yao[1], Chenrui Jing[1, 2], Ziting Li[1, 2], and Ya Cheng[1, *]**

[1]State Key Laboratory of High Field Laser Physics, Shanghai Institute of Optics and Fine Mechanics, Chinese Academy of Sciences, Shanghai 201800, China

[2]University of Chinese Academy of Science, Beijing 100049, China

Email: chuwei0818@qq.com; ya.cheng@siom.ac.cn.

**Abstract**. We experimentally investigate generation of backward 357 nm $N_2$ laser in a gas mixture of $N_2$/Ar using 800-nm femtosecond laser pulses, and examine the involved gain dynamics based on pump-probe measurements. Our findings show that a minimum lifetime of population inversion in the excited $N_2$ molecules is required for generating intense backward nitrogen lasers, which is ~0.8 ns under our experimental conditions. The results shed new light on the mechanism for generating intense backward lasers from ambient air, which are highly in demand for high sensitivity remote atmospheric sensing application.

## 1. Introduction

Atmospheric remote sensing based on laser techniques has been emerging as innovative tools for applications in air pollution monitoring and detection of hazardous chemicals [1-3]. To facilitate highly sensitive remote sensing with single-sided detection scheme - a geometry that inherently allows for maximum flexibility in terms of areas to be covered, a directional backward (BW) coherent pump source which can be directly generated at the detection site is in great demand. Recently, the potential to accomplish this seemingly unrealistic goal has been demonstrated by generating free-space remote atmospheric lasers using air molecules as the gain medium [4-7].

So far, the demonstrated air lasers can be categorized into two types. In the first type of air lasers, population inversion is achieved by dissociation of molecular oxygen or nitrogen followed by excitation of the atomic fragments using high-peak-intensity picosecond ultraviolet (UV) lasers, which gives rise to bidirectional amplified spontaneous emissions (ASEs) at either 845 nm from oxygen atoms or 870 nm from nitrogen atoms [8-10]. The second type of air lasers are realized by focusing intense ultrafast laser pulses in air which created population inversion conditions either in neutral $N_2$ molecules or in $N_2$ ions [11-15]. Although the pump mechanism behind the air laser from neutral $N_2$ molecules has been clarified which can be attributed to electron collisional excitation, the mechanism of the air laser from $N_2$ ions is still far from being fully understood and under intensive investigation.

Interestingly, although significant population inversion has been generated in both the two types of air lasers mentioned above, intense backward ASE signals can only be observed in the former type of lasers, namely, the atomic air lasers realized with picosecond ultraviolet pump lasers. A careful examination shows that although high gain coefficients can be observed in the forward (FW) direction for both the atomic and molecular air lasers, they have distinctly different gain dynamics. Noticeably, the lifetimes of the population inversions in the atomic and molecular air lasers are on the nanosecond and picosecond levels, respectively [10, 14]. Quantitative investigation on the generation of ASE-type air lasers under variable population inversion lifetime has not been performed until now. Here, we present direct evidence of the role of population inversion lifetime on the generation of backward air laser from $N_2$ molecules. Specifically, we adopt a technique developed by Kartashov *et al.* to achieve sufficiently long population inversion lifetimes (on the order of nanosecond) in mixtures of $N_2$ and Ar gases at different concentration ratios [6]. To enable tuning of the lifetime of population inversion, we develop a dual-pulse pump scheme. In the scheme, the first pulse plays the role of establishing the population inversion between the excited states $C^3\Pi_u$ and $B^3\Pi_g$ of $N_2$ molecules, as have been reported in Refs. [6, 14]. In contrast, the time-delayed second pulse plays a role of terminating the population inversion by depleting the excited $N_2$ molecules through photoionization. This scheme allows tuning of the population inversion lifetime by varying the time delay between the pump and probe pulses. Our result provides quantitative information on the population inversion lifetime required for generating backward air lasers from neutral $N_2$ molecules excited by intense femtosecond laser pulses.

The remainder of the paper is organized as follows. In section 2, we investigate the gain dynamics of lasing actions in $N_2$/Ar mixture excited by intense femtosecond laser pulses. Based on the results, we optimize the experimental conditions including the focal conditions of femtosecond laser and the pressures of $N_2$ and Ar gases for maximizing the lasing signal. In section 3, we perform systematic investigation on the dependence of backward lasing signal on the lifetime of the population inversion in $N_2$ molecules. In section 4, we summarize the major results and discuss the implication of our work on the future development of backward laser in remote air.

## 2. Gain dynamics of lasing actions in $N_2$/Ar mixture

*2.1 Experimental setup*

The pump-probe experimental setup for investigating the gain dynamics of air lasers generated in a gas mixture of $N_2$/Ar is schematically illustrated in Fig. 1, which is similar to that used in our previous works [16-17]. Briefly, femtosecond laser pulses (1 kHz, 800 nm, ~40fs) from a commercial Ti: sapphire laser system (Legend Elite Cryo PA, Coherent, Inc.) were divided into two paths with an 80/20 beam splitter (BS). One beam with a pulse energy of ~10.5 mJ served as the pump and the other with a pulse energy of ~2 mJ was used to produce wavelength-tunable seed pulses. In our case, the pulse with energy of 2 mJ was first reduced to ~3 mm in diameter by a telescope system and its spectrum was broadened using a piece of 20-mm-long BK7 glass. Then, the 800 nm pulse was frequency doubled using a 2-mm-thickness $\beta$-barium borate crystal (BBO), which acted as the seed pulse to generate seed-amplification-based air laser emissions. The center wavelength of seed pulse can be continuously tuned in a broad range around 357-nm by rotating the angle of BBO crystal. Both the pump and seed pulses were linearly polarized and their polarization directions were perpendicular to each other. The time delay between the pump and seed pulse was controlled by a motorized linear translation stage with a temporal resolution of ~16.7 fs. After being combined by a dichroic mirror (with high reflectivity at ~800 nm and high transmission at ~400 nm), the pump and seed pulses were collinearly focused into a gas chamber filled with $N_2$/Ar mixture by a 150-cm focal-length lens. After interaction with the gas mixture in the chamber, the exiting pulses were first collimated by a 150-cm focal-length lens and then passed through two dichroic mirrors to remove the intense 800 nm pulses. Furthermore, to remove the strong supercontinuum white light generated during the propagation of the intense pump pulses in the gas chamber, a Glan-Taylor prism was used to completely separate the pump beam and 357-nm $N_2$ laser because of their different polarization directions. Lastly, the forward spectra of the $N_2$ laser were recorded using a 1200-grooves/mm grating spectrometer (Andor Shamrock 303i) which has a spectral resolution of ~0.06 nm.

*2.2 Results and discussion*

Figure 2(a) shows a typical forward 357 nm ($C^3\Pi_u(\nu=0) \to B^3\Pi_g(\nu'=1)$ of $N_2$) laser spectrum (blue solid curve) generated by focusing both the pump and seed pulses into the gas chamber which was filled with a mixture of 300-mbar $N_2$ gas and 900-mbar Ar gas. For comparison, both the spectra of the injected seed pulse and of the ASE laser at 357 nm wavelength produced by only the pump pulses are plotted with the red dashed and black dash-dotted curves, respectively. The inset in Fig. 2(a) clearly indicates that the 357-nm $N_2$ laser generated with the seed pulse has a nearly perfect linear polarization. Similar to the previously reported seed-amplification-based air lasers, the polarization direction was again measured to be parallel to that of the seed pulses [14]. Meanwhile, we also recorded the backward 357 nm ASE laser spectrum generated with only the pump pulses, as shown in Fig. 2(b).

In our experiments, we found that the $N_2$ laser signal sensitively depends on the gas pressures. In particular, under the conditions of our pump laser and focal system, the strongest laser signal was observed at the gas pressures of 300 and 900 mbar of $N_2$ and Ar gases, respectively. Figure 3(a) shows the dependences of the forward (blue squares curve) and backward (red triangles curve) 357 nm laser signals on the Ar gas pressure. The reason that we mainly investigated the laser signal as a function of Ar gas pressure is that the concentration of Ar gas can strongly influence the lifetime of population inversion in the $N_2$ molecules as a consequence of the excitation mechanism of $N_2$ laser [18]. This feature has also been confirmed by our observations. It can be seen that the lasing action is initiated when the Ar gas pressure is increased to ~550 mbar. Then, the laser signal first grows rapidly with the gas pressure until the Ar gas pressure reaches ~900 mbar. Afterwards, the laser signal drops with the gas pressure and eventually vanishes at an Ar gas pressure of ~1600 mbar. We speculate that the quenching of the laser signal should be due to depletion of the excited $N_2$ molecules through a collisional quenching reaction $N_2(C^3\Pi_u) + Ar \rightarrow N_2(B^3\Pi_g) + Ar$ which becomes more severe at the Ar gas pressures higher than ~900 mbar [19].

Figure 3(b) shows the normalized intensity of 357 nm $N_2$ laser as a function of the time delay between the pump and seed pulses measured at a fixed gas pressure of 300 mbar of $N_2$ gas and varied pressures of 600, 900, and 1200 mbar of Ar gas. At zero time delay, the relative intensities of the $N_2$ laser measured under these gas conditions can be evaluated from the data labeled with A, B, and C in Fig. 3(a). Here, the zero time delay between the pump and seed pulses was determined by observing the beginning of plasma defocusing of the seed pulses induced by the pump pulses, and the positive delay means that the seed pulse is behind the pump pulse. It can be seen that at all the gas pressures of Ar, lasing actions (i.e., amplification of the seed pulses) occurred at time delay of ~3 ns between the pump and seed pulses. We roughly evaluated the population inversion lifetimes (full width at half maximum, FWHM) at the Ar gas pressures of 600 mbar, 900 mbar, and 1200 mbar with the least-square fitting curves as shown in Fig. 3(b), which are ~3 ns, ~3.4 ns, and ~2.3 ns, respectively. The longest population inversion lifetime is observed at the Ar gas pressure of 900 mbar, which is consistent with the strongest air laser signal generated under this condition.

The results in Fig. 3 indicate that pumping $N_2$ molecules with excited metastable Ar atoms can lead to much longer population inversion lifetime than that achievable with the electron collisional excitation in pure $N_2$ gas, even intense, ultrashort laser pulses are used in both cases [14, 18]. The difference in turn leads to essentially different lasing behaviors, i.e., the former scheme has successfully generated strong ASE $N_2$ laser in both forward and backward directions, whereas with the latter scheme only forward laser signals can be generated with a decent pulse energy. To quantitatively reveal the role of the population inversion lifetime, we design the following experiment to generate air lasers with variable population inversion lifetime of the nitrogen molecules, and investigate how the backward lasing behavior relies on it.

## 3. Dependence of backward lasing on the population inversion lifetime

*3.1 Experimental setup*

The experimental setup is schematically shown in Fig. 4, in which the same femtosecond laser system was employed. Again, the femtosecond laser beam was split into two for generating a dual-pulse source. The first pulse was used as the pump pulse for generating the population inversion in $N_2$ molecules which can enable generation of the backward ASE laser. The time-delayed second pulse was used to quench the lasing action, which is referred to as the quenching pulse in the following. The center wavelength of both pulses was ~800 nm. The energies of pump and quenching pulses were measured to be ~4.2 mJ and ~0.35 mJ before the focal lens, respectively. The peak intensity of quenching pulses is significantly lower than that of the pump pulses because for the $N_2$ molecules at excited states, their ionization potentials are much lower than that of the $N_2$ molecules at the ground state. Both the pump and quenching pulses were linearly polarized and their polarization directions were parallel to each other.

The time delay between the two pulses was controlled using a motorized linear translation stage. After combined by a beam splitter with 70% reflectivity and 30% transmission at 800 nm, the pump and quenching pulses were collinearly focused by a 150-cm focal-length lens into the chamber filled with the gas mixture of $N_2$/Ar. It should be mentioned that using such a beam combining scheme, energy loss of light is inevitable. However, since our pump pulses were sufficiently energetic, both forward and backward laser signals were generated using the illustrated dual-pulse pump scheme. We inserted a dichroic mirror (with high reflectivity at ~400 nm and high transmission at ~800 nm) in the probe beam path to extract the backward $N_2$ laser emissions. The same grating-based spectrometer was used to record the spectra of the backward laser at 357 nm wavelength.

*3.2 Results and discussion*

Figure 5(a) shows the 357 nm $N_2$ laser spectra measured in the backward direction with (red dash-dotted curve) and without (blue solid curve) the second pump pulse. In this specific case, the time delay between the pump and quenching pulses was set at ~2.3 ns when a complete quenching of the backward laser was observed. The gas pressures of $N_2$ and Ar were 300 mbar and 900 mbar, respectively.

In Fig. 5(b), we present the measured intensity of the backward $N_2$ laser signal as a function of the time delay between the pump and quenching pulses (red dashed curve). For comparison, the forward gain dynamics in the $N_2$ molecules measured under the same pump conditions was indicated by the blue dotted curve. The two curves together reveal the crucial role played by the population inversion lifetime on the generation of backward $N_2$ ASE laser. From the blue dotted curve showing the gain dynamics, one can see that the population inversion was established ~3.2 ns after the pump pulse. Besides, it was found in the red dashed curve that when the time delay between pump and quenching pulses reached ~4 ns, the backward laser signal started to appear. This observation implies that when the quenching pulse arrives earlier (i.e., in a time window between ~3.2 ns and ~4 ns behind the pump pulse), the population inversion will be destroyed before the ASE could be effectively established in the population inverted $N_2$ molecules. We determine that the difference between the above two times, which is ~0.8 ns, indicates the minimum population inversion lifetime required for generating the backward ASE air laser.

To further confirm the above physical understanding, we performed the same measurement at a different gas pressure of $N_2$ (500 mbar), as shown in Fig. 5(c). We found that under the gas condition, the population inversion was established ~2.1 ns after the arrival of pump pulse and the backward laser signal started to appear when the time delay between pump and quenching pulses reached ~2.9 ns. Therefore, the minimum population inversion lifetime required for generating the backward ASE air laser under this condition is estimated to be ~0.8 ns, which is very close to the result as indicated by Fig. 5(b). Moreover, both Fig. 5(b) and 5(c) show that for a sufficiently long time delays between the pump and quenching pulses, the backward laser signals grow with the increasing time delay, and finally become saturated. The saturation is simply due to the fact that when the quenching pulse arrives too late, i.e., after the generation of backward ASE $N_2$ laser has been completed, the quenching pulse will no longer be able to disturb the backward ASE signal.

It should be noted that the population inversion lifetimes measured in pure $N_2$ gas excited by ultrafast laser pulses are typically one order of magnitude shorter (i.e., a few tens of picoseconds), which are sufficient for generating forward ASE lasers but insufficient for generating an intense backward ASE laser [11, 14]. Our result suggests that for generating backward ASE air laser from $N_2$ molecules, the minimum population inversion lifetime should be on a timescale of ~1 ns.

## 4. Summary and outlook

In summary, we have investigated generation of backward ASE $N_2$ lasers in $N_2$/Ar mixture with variable gas pressures and the corresponding gain dynamics. Based on the dual-pulse pumping scheme, we achieved tuning of the population inversion lifetime in $N_2$ molecules, and studied the dependence of backward air laser signal on the population inversion lifetime. Our findings suggest that a minimum population inversion lifetime of ~0.8 ns is required for generating backward ASE air lasers in femtosecond laser induced air plasma. The insight gained from the current investigation indicates that future effort in generation of the backward air lasers from strong-field-excited $N_2$ molecules or $N_2$ molecular ions should focus on the development of innovative approaches for extending the lifetimes of population inversion.

**Acknowledgements**

This work is supported by the National Basic Research Program of China (Grants No 2011CB808100 and No. 2014CB921300), the National Natural Science Foundation of China (Grants No. 11127901, No. 11134010, No. 11204332, No. 11304330, and No. 11404357), and the Shanghai "Yang Fan" program (Grant No. 14YF1406100).

**Captions of figures:**

Figure 1. Schematic of the experimental setup. HR: high reflectivity mirror; DM: dichroic mirror; GT: Glan-Taylor prism.

Figure 2. (a) Typical forward (FW) lasing spectra obtained with (blue solid) and without (black dash-dotted) the seed pulses. Inset: polarization property of 357-nm nitrogen laser generated with a seed pulse. (b) Typical backward (BW) lasing spectrum generated obtained after the seed pulses are blocked.

Figure 3. (a) The dependence of forward (blue squares) and backward (red triangles) nitrogen laser signal on Ar gas pressure when $N_2$ gas pressure is fixed at ~300 mbar. (b) The temporal evolution of the 357-nm nitrogen laser with the time delay between the pump and seed pulses at the Ar gas pressure of 600 (red circles), 900 (black squares), and 1200 mbar (blue triangles), respectively. The red solid, black dotted, and blue dashed curves are the least-square fitting curves of the data obtained at 600, 900, and 1200 mbar, respectively, based on which the population inversion lifetimes are estimated.

Figure 4. Schematic of the dual-pulse pumping experimental setup. HR: high reflectivity mirror; BS: beam splitter; DM: dichroic mirror.

Figure 5. (a) A typical 357-nm nitrogen laser spectra measured in the backward direction with (red dash-dotted curve) and without (blue solid curve) the quenching pulse. (b) The backward ASE-based nitrogen laser signal (red dashed curve) as a function of the time delay between the pump and quenching pulses. The pressures of $N_2$ and Ar gases are 300 and 900 mbar, respectively. The gain dynamics (blue dotted curve) is shown for comparison. (c) The same measurement as in (b) but at a different $N_2$ gas pressure of 500 mbar.

Fig. 1

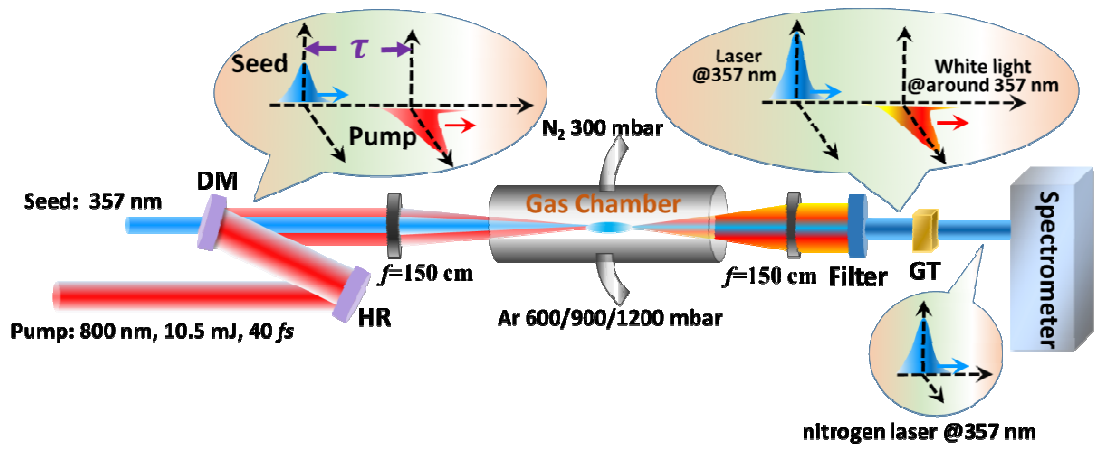

Fig. 2

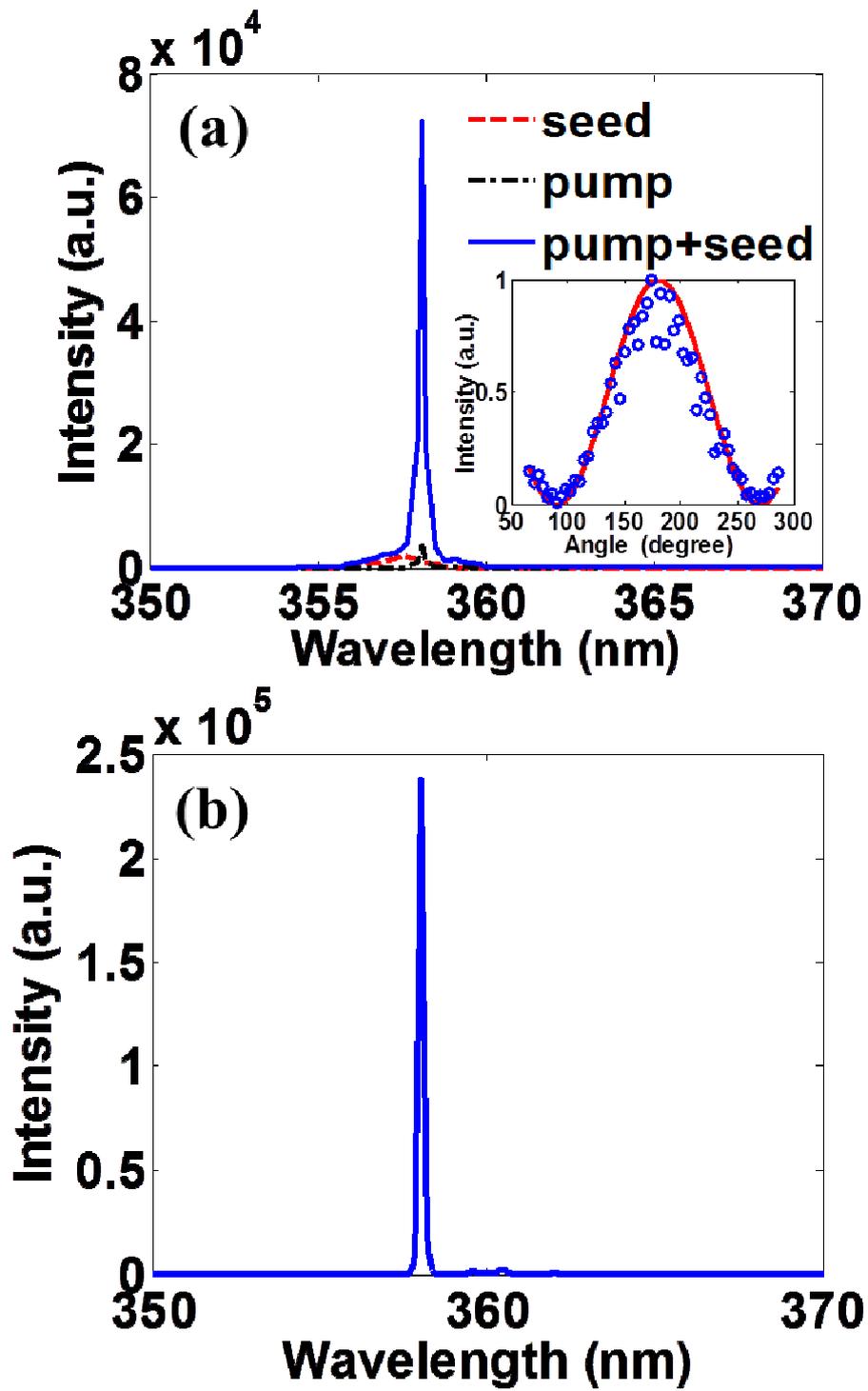

Fig. 3

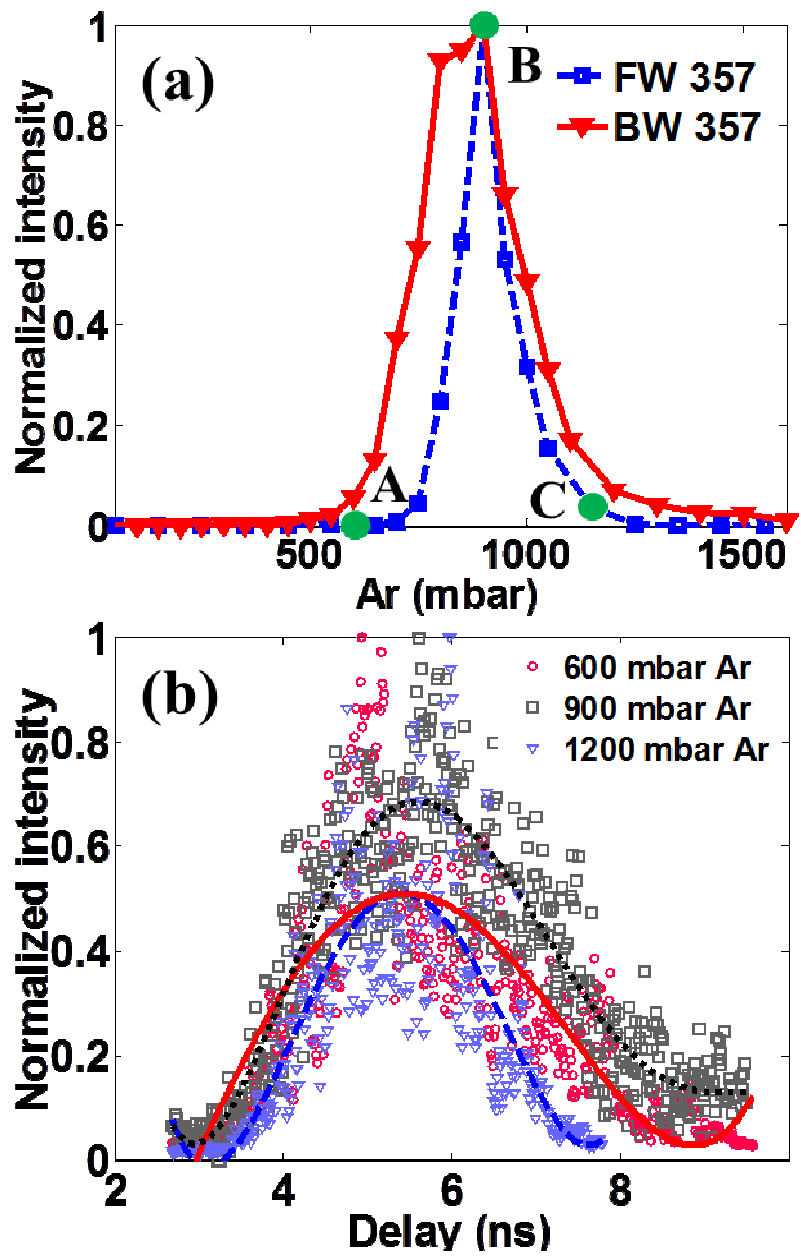

Fig. 4

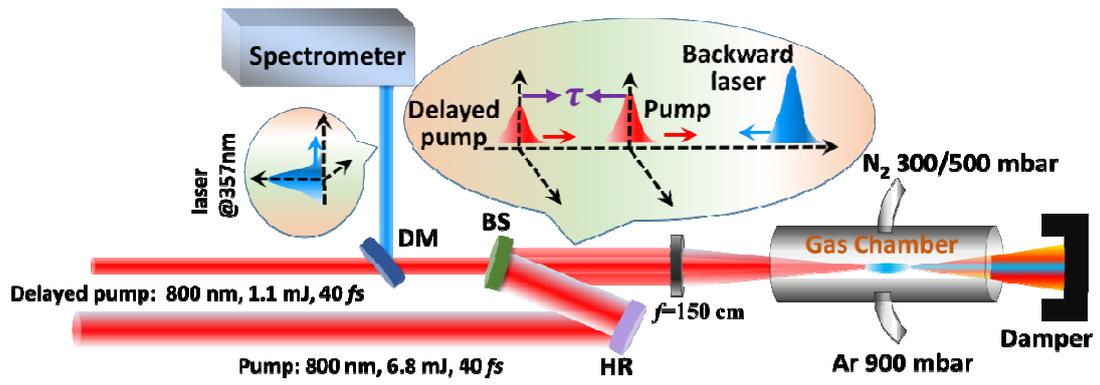

Fig. 5

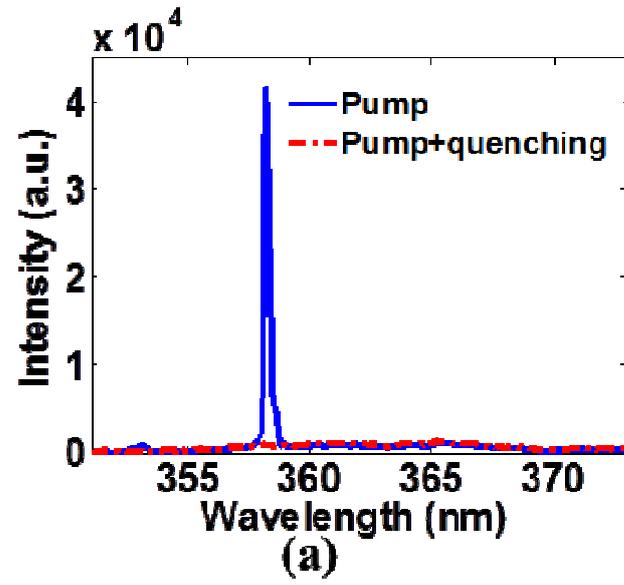

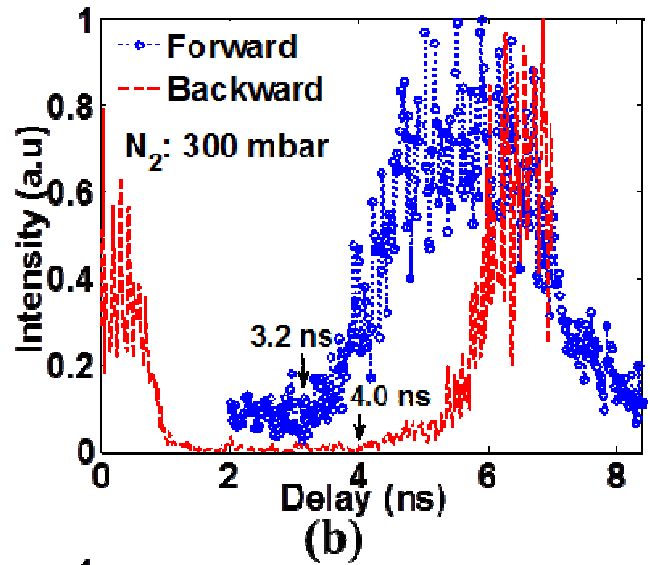

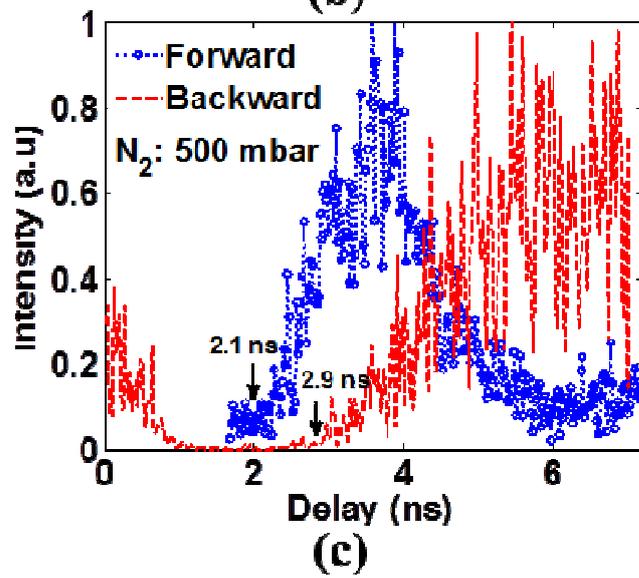